\documentclass[aps,prl,showpacs,amsmath,nofootinbib,
superscriptaddress,twocolumn,preprintnumbers]{revtex4}
\usepackage{natbib}
\usepackage{amsmath}
\usepackage{amssymb}
\usepackage{graphicx}
\usepackage{color}
\usepackage[makeroom]{cancel}
\usepackage[normalem]{ulem}

\begin{document}
\title{A Globally Unevolving Universe}
\author{Meir Shimon}
\affiliation{School of Physics and Astronomy, 
Tel Aviv University, Tel Aviv 69978, Israel}
\email{meirs@wise.tau.ac.il}
\begin{abstract}
A scalar-tensor theory of gravity is formulated in which $G$ and particle masses 
are allowed to vary. The theory yields a globally static cosmological model with 
no evolutionary timescales, no cosmological coincidences, and no flatness and 
horizon `problems'. It can be shown that the energy densities of dark energy 
($\rho_{DE}$) and non-relativistic baryons and 
dark matter ($\rho_{M}$) are related by $\rho_{DE}=2\rho_{M}$, in agreement 
with current observations, if DE is associated with the canonical kinetic and potential 
energy densities of the scalar fields. 
Under general assumptions, the model favors {\it light} fermionic 
dark matter candidates (e.g., sterile neutrinos).
The main observed features of the CMB are 
naturally explained in this model, including the spectral flatness of its perturbations 
on the largest angular scales, and the observed 
adiabatic and gaussian nature of density perturbations. 
More generally, we show that many of the cosmological observables, 
normally attributed to the dynamics of expanding space, could be 
of kinematic origin. In gravitationally bound systems, 
the values of G and particle masses spontaneously freeze out by 
a symmetry breaking of the underlying conformal symmetry, 
and the theory reduces to standard general relativity (with, e.g., all 
solar system tests satisfied).
\end{abstract}
\maketitle

{\bf Introduction}.- 
In standard cosmology physical processes are regulated by space expansion, 
i.e. the time-dependent Hubble scale provides the `clock' for the evolving 
temperature and density of matter and radiation, resulting in a sequence 
of cosmological epochs. This clock is only meaningful if other time scales, 
e.g. the Planck time, or the Compton scale, evolve differently. 

In the current standard (cosmological) model (SM) the conjecture is made 
that a brief inflationary phase in the very early universe provides the 
theoretical framework for resolving the `puzzles' of observed spatial 
flatness, deduced super-horizon correlations in the cosmic microwave 
background (CMB) radiation, and absence of primordial cosmic defects [1-3]. 
Generic inflationary models seem to be able to explain 
the nearly flat power spectrum 
of density perturbations, their gaussian nature, and adiabatic initial conditions. 
In addition, they generically predict a unique imprint on the polarization state of 
the CMB generated by inflationary-induced gravitational waves [4-6].

Yet, a few puzzling features remain unexplained; baryons account for only 
$\approx 5\%$ of the energy budget of the universe, with the rest in some form 
of dark energy (DE) and dark matter (DM). The latter has so far evaded detection 
by terrestrial experiments, and the former does not seem to cluster on sub-horizon 
scales. The nature of DM is yet unknown, and according to standard lore DE is 
very likely described by a scalar field with an equation of state (EOS) which very 
closely mimics vacuum energy, but with an amplitude $O(10^{122})$ smaller than 
naively expected for a cosmological constant [7-9]. 

Although DE and nonrelativistic (NR) matter evolve very differently in the SM, 
the current energy density of the former is about 2/3 of the total energy budget of the 
universe [10, 11]. Common interpretation of cosmological observations is that the 
transition from matter-dominated to DE-dominated expansion took place only recently. 
This provoked anthropic considerations [12-15] to explain the uniqueness of the current 
epoch. Other suggestions to explain the small value of DE include models of dynamic 
DE [16-19], holographic cosmology [20-24], K-essence [25, 26], modified gravity [27, 28], 
etc. Yet, it is fair to say that none has gained consensus. These and other puzzles, 
e.g. [29, 30], have been contemplated for quite some time now; generally, they are 
viewed as either coincidences of nature, or as possible indications of the need 
for new physics.

It should be stressed from the outset that {\it the only cosmological observables are 
dimensionless quantities on the past light cone}. From this perspective, cosmological 
models are practically redundant in the sense that they typically describe the evolution 
of dimensional quantities in the entire spacetime, and consequently most of the rich 
dynamics characterizing these models is not readily amenable to comparison with 
observations.

We show that the above cosmological puzzles may be attributed to our 
conventional system of units which is based on constant {\it dimensional} quantities, 
namely, natural units, such as Planck length, and Compton wavelength. In contrast to other 
scalar-tensor theories, e.g. [31-35], we propose a `single-clock' cosmological model 
characterized by Minkowski background metric with no global background evolution. 
By projecting certain dimensionless quantities derived from the model on the past 
light cone a few well known SM results are reproduced, and new insight is gained. 
All this is achieved in a framework that does not recourse to GUT scale physics, which is 
many orders of magnitudes beyond our current experimental reach. Rather, the proposed 
framework replaces the {\it dynamical} SM with a much simpler {\it kinematic} 
model, the latter is only required to satisfy 
self-consistent boundary conditions on the past light cone -- essentially the observable 
universe -- thereby explaining, e.g., the observed ratio of DE and NR 
matter, and estimating the mass of a fermionic cold 
DM (CDM) candidate to be likely below $1GeV$. 

{\bf Theoretical Framework}.-
A scalar-tensor theory of gravity, linear in the curvature scalar, $R$, 
can be summarized by the following action given in $\hbar$ units [36, 37]
\begin{eqnarray}
S/\hbar&=&\int\left[\frac{1}{2}F(\varphi^{K})R
-\frac{1}{2}g_{IJ}(\varphi^{K})\varphi_{\mu}^{I}\varphi^{J,\mu}
-V(\varphi^{K})\right.\nonumber\\
&+&\left.\mathcal{L}_{M}(\varphi^{K})/(\hbar c)\right]\times\sqrt{-g}d^{4}x
\end{eqnarray}
where the integration measure is $d^{4}x=c dt\cdot d^{3}x$,
summation convention is implied on both greek and capital Latin letters, 
the $N$ scalar fields $\varphi^{K}$ are labeled by $I, J, K=1, 2, ...., N$, 
and $\hbar$ and $c$ have their usual meaning. The potential $V$ is an explicit 
function of the scalar fields. In general, the matter Lagrangian $\mathcal{L}_{M}$ 
accounts for the entire mass-energy contributions to the energy density 
of the universe, i.e. DE, DM, baryons, electrons, neutrinos, and radiation. In 
addition to the spacetime metric $g_{\mu\nu}$ we introduce the dimensionless 
$g_{IJ}$ which is a `metric in field space'. $R$ is 
calculated from $g_{\mu\nu}$, and $f_{\mu}\equiv f_{,\mu}$ is the derivative 
of a function $f$. We allow $\mathcal{L}_{M}$ to explicitly depend on $\varphi^{K}$.

The Einstein-Hilbert (EH) action of conventional general relativity (GR)
is recovered by setting all scalar fields to constant values. Consistency with 
GR requires $F(\varphi^{K})\equiv\kappa^{-1}$ and $V=\Lambda/\kappa$, 
where $\kappa\equiv 8\pi G\hbar/c^{3}=O(l_{P}^{2})$, and $G$, $l_{P}$, and $\Lambda$ 
are the gravitational constant, Planck length, and cosmological constant, 
respectively. 
Eq. (1) describes GR in dynamic units, with $\varphi^{k}$ non-minimally coupled to gravitation. 
In general, scalar fields can make both negative and positive contributions to the 
energy density; while the former can replace the gravitational energy density, 
thereby `offset' the standard expansion, the latter can be identified with DE, 
in which case $\mathcal{L}_{M}$ accounts only for the remaining 
contributions to the total energy density.

Applied to a homogeneous and isotropic background, this theory admits 
classical solutions characterized by a universal evolution of all quantities 
with the same physical units, namely, global dimensionless ratios are 
fixed, indeed a cornerstone of the approach adopted here. In this special 
case the theory reduces to a single-scalar model where 
$\varphi^{I}\equiv\lambda^{I}\varphi$, with dimensionless coupling constants 
$\lambda^{I}$.
For $V=\lambda\varphi^{4}$ and $F\equiv\beta\varphi^{2}$,  
where $\beta<0$ and $\lambda>0$, the EH action is recovered from Eq. (1) 
by setting $\varphi^{2}=(8\beta\pi G)^{-1}$ and $\lambda=8\pi\beta^{2}G\Lambda$.

The field equations derived from variation of Eq. (1) with respect to $g_{\mu\nu}$, 
and $\varphi$ are [36, 37]
\begin{eqnarray}
F\cdot G_{\mu}^{\nu}&=&T_{M,\mu}^{\nu}+\left(\varphi_{\mu}\varphi^{\nu}
-\frac{1}{2}\delta_{\mu}^{\nu}\varphi^{\alpha}\varphi_{\alpha}\right)\nonumber\\
&-&\delta_{\mu}^{\nu}V+F_{\mu}^{\nu}-\delta_{\mu}^{\nu}F_{\alpha}^{\alpha}\nonumber\\
\Box\varphi&+&\frac{1}{2}F^{,\varphi}R-V^{\varphi}=-\mathcal{L}_{M}^{;\varphi}\nonumber\\
T_{M,\mu;\nu}^{\nu}&=&\mathcal{L}_{M,\varphi}\varphi_{\mu}
\end{eqnarray}
where $G_{\mu}^{\nu}$ is Einstein's tensor, $g_{IJ}\lambda^{I}\lambda^{J}\equiv 1$ and 
$f_{\mu}^{\nu}\equiv[(f)_{,\mu}]^{;\nu}$, with $f_{;\mu}$ denoting covariant 
derivatives of $f$, $\Box f$ is the covariant Laplacian, 
$(T_{M})_{\mu\nu}\equiv\frac{2}{\sqrt{-g}}\frac{\delta(\sqrt{-g}\mathcal{L}_{M})}
{\delta g^{\mu\nu}}$, and the last of Eqs. (2) implies that energy-momentum is not 
generally conserved. Indeed, energy-momentum is clearly not conserved when $G$, 
$\Lambda$, and particle masses, are time-dependent.

{\bf Homogeneous and Isotropic Spacetimes}.-
The line element describing a homogeneous and isotropic spacetime in cosmic 
coordinates is
\begin{eqnarray}
ds^{2}=-(cdt)^{2}+a^{2}\left(\frac{dr^{2}}{1-Kr^{2}}+r^{2}d\Omega\right),
\end{eqnarray}
where $d\Omega\equiv d\theta^{2}+\sin^{2}\theta d\phi^{2}$ is 
a differential solid angle, $a(t)$ is a scale factor, and 
$K$ is a spatial curvature constant.
Eqs. (2) are then
\begin{eqnarray}
3F(H^{2}+\frac{K}{a^{2}})&=&\rho_{M}+\frac{1}{2}\dot{\varphi}^{2}+V-3H\dot{F}\nonumber\\
-2F(\dot{H}-\frac{K}{a^{2}})&=&\rho_{M}(1+w_{M})+\dot{\varphi}^{2}+\ddot{F}-H\dot{F}\nonumber\\
\ddot{\varphi}+3H\dot{\varphi}&-&\frac{1}{2}F^{,\varphi}R+V^{\varphi}=\mathcal{L}_{M}^{,\varphi} ,
\end{eqnarray}
where $R=6(2H^{2}+\dot{H}+\frac{Kc^{2}}{a^{2}})$, $H\equiv\dot{a}/a$, and 
$(T_{M})_{\mu}^{\nu}=\rho_{M}\cdot diag(-1,w_{M},w_{M},w_{M})$, 
i.e. there are no shear and momentum-flow in a homogeneous background, and 
$w_{M}\equiv P_{M}/\rho_{M}$ is the EOS of matter.

A homogeneous and isotropic model (with $\rho_{M}=-\mathcal{L}_{M}>0$) 
admits nonvanishing constant $\varphi$, only when $\beta R\varphi^{2}>0$, by virtue of a 
symmetry breaking of the underlying conformal symmetry where $\beta R\varphi^{2}/2$ plays 
the role of a mass term. 
For $\beta R\varphi^{2}\leq 0$, i.e. $R\leq 0$, only dynamical $\varphi$ is a nontrivial solution.
Combining the traces of the first two of Eqs. (2) results in $\mathcal{L}_{M}=T$.
A solution, of the form $\dot{a}^{2}+Kc^{2}=0$ (that amounts to 
$G_{0}^{0}=G_{i}^{i}=R=0$, i.e. flat spacetime), with $\varphi$ scaling as $\varphi=\varphi_{0}/a$ 
(with subscript `0' denoting present values here and throughout) results in the following constraints
\begin{eqnarray}
&&\rho_{M}+H^{2}(\frac{1}{2}+6\beta)\varphi^{2}+V=0\nonumber\\
&&\rho_{M}(1+w_{M})+H^{2}(1+8\beta)\varphi^{2}=0\nonumber\\
&&-H^{2}\varphi+V^{\varphi}=\mathcal{L}_{M}^{\varphi},
\end{eqnarray}
where $\mathcal{L}_{M}^{\varphi}\equiv\mathcal{L}_{M,\varphi}$ 
and $V^{\varphi}\equiv V_{,\varphi}$. Under the condition that the energy density 
associated with the i'th species is $\rho_{i}\propto a^{-4}$, 
(irrespective of EOS), 
and therefore $\rho_{M}\propto a^{-4}$ and $\mathcal{L}_{M}\propto a^{-4}$, the last of 
Eqs. (5) becomes
\begin{eqnarray}
\rho_{M}(-1+3w_{M})=\mathcal{L}_{M,\varphi}\varphi.
\end{eqnarray}
This condition implies that there is essentially no global evolution, and in case of a 
NR matter $\mathcal{L}_{M}=-\rho_{M}$ if $\rho_{M}\propto\varphi$. 
In fact, combining the derivative of the first of Eqs. (5) with the last 
is only consistent if $\beta=-1/6$ as is expected from our fully conformal model [38-40].

A tantalizing possibility is that the (canonical) kinematic and potential energies associated with the 
scalar fields may account for DE, and the scalar fields themselves replace the masses of 
NR baryons and fermionic DM particles with $w_{M}\approx 0$. From the first two of 
Eqs. (5), we see that 
\begin{eqnarray}
\rho_{DE}=2\rho_{M}
\end{eqnarray}
where $\rho_{M}$ stands for the NR (DM and baryonic) matter. 
Remarkably, this relation conforms well with best fit estimates from SM, e.g. [10, 11]. 
In the model explored here the ratio $\rho_{DE}/\rho_{M}$ is fixed and there is no 
`coincidence problem' in the current epoch. The apparent coincidence in the SM stems from 
arbitrarily fixing dimensional units; doing so at any past or future time would have resulted 
in the same $\rho_{DE}/\rho_{M}\approx 2$. From the above discussion it is clear 
that $-K=O(\Lambda)=O(H_{0})$. In a {\it stationary} universe, unlike in the SM, 
the fact that the curvature radius is not much larger than the Hubble radius has no direct 
observational consequences since there is no a-priori expectation for the coherence 
scale of CMB perturbations, i.e. $O(1^{\circ})$ does not imply $K=0$.

In our model, the EOS of DE and NR matter are 
$-1/3$ and $0$, respectively, in contrast with $-1$ and $0$ in the SM. 
However, in the SM the underlying Lagrangian of DE is unknown 
and $w_{DE}=-1$ was empirically deduced from observations based on GR 
which relies on a unit convention that conserves energy-momentum. In other words, 
$w_{DE}$ is not an observable. It is therefore not surprising that in other conventions, such 
as the one adopted here, the EOS might be different. 

We note that $a=\sqrt{-K}t$ is Milne's solution to empty spacetime 
in cosmic coordinates [41-44]. 
Unlike Milne's original work that was based on special relativity 
and therefore implicitly assumed that $l_{P}$, 
$l_{\Lambda}$ ($\Lambda\equiv 12 l_{\Lambda}^{-2}$), 
and $l_{C}=\hbar/(mc)$ are constant, the framework proposed here accounts for the 
entire energy density in the universe.

{\bf Kinematics of the Observed Background Universe}.-
Incoming radial null geodesics in Milne spacetime (Eq. 3 with $a=\sqrt{-K}t$) are $\sinh\eta=-\sqrt{-K}r$, 
where $\eta\equiv\ln(t)$ is conformal time. Setting $\xi\equiv -\eta$ where $\xi$ is the 
{\it local} `rapidity parameter', the dimensionless `velocity' and 'Lorentz factor' are, respectively, 
$\beta\equiv v/c\equiv\tanh\xi=\sqrt{-K}r/\sqrt{1-Kr^{2}}$, and $\gamma\equiv\cosh\xi=(1-Kr^{2})^{1/2}$. 
Recasting Milne spacetime in `light-cone conformal coordinates' 
\begin{eqnarray}
t'=t\cosh\xi, \, \, \, r'=ct\sinh\xi,
\end{eqnarray} 
energy densities now scale as $t^{-4}=(t'^{2}-r'^{2})^{-2}$, i.e. the latter are inhomogeneous 
in the new frame, which is in fact the Minkowski frame. Note that this transformation is singular at $t=0$, 
essentially the natural choice for the `origin' of (cosmic) time. Since the coordinate transformation of Eq. (8) 
mixes time and radial coordinates the energy-momentum tensor in Minkowski 
coordinates includes new off-diagonal terms that represent radial momentum flow in this frame. 
This {\it is} the origin of the cosmological redshift 
of radiation emitted at a (Minkowski) spacetime point as measured by an observer at the origin. 
Specifically, in Minkowski coordinates $T_{r'}^{t'}=(1-Kr^{2})^{1/2}\sqrt{-K}r(\rho+P)$. 
In the limit $-Kr^{2}\ll 1$, $\beta\approx\sqrt{-K}r$, 
which is perceived by the observer as recession of the distant emitter with apparent velocity 
$v=H_{0}r$, i.e., the Hubble law with $H_{0}=c\sqrt{-K}$. Transforming to the appropriate Minkowski 
coordinates it is then found that $v\approx H'r'$ where $H'=t'^{-1}$, i.e. $t'=H_{0}^{-1}$ at present.
As a result 
$t/t_{0}=e^{\eta}=\sqrt{(1+\beta)/(1-\beta)}$, i.e. the frequency redshifts 
$\propto\sqrt{(1+\beta)/(1-\beta)}$ where the velocity monotonically increases 
with $r$. Similarly, energy densities scale 
as $\rho\propto[(1+\beta)/(1-\beta)]^{2}$. 
Since by definition $\rho\equiv\rho_{0}(1+z)^{4}$, 
this implies that the distance-redshift relation reads $\sqrt{-K}r=z(1+z/2)/(1+z)$. Indeed, 
this differs from the corresponding SM relation; however, cosmological 
(e.g. luminosity, angular diameter) distances 
are not observables, rather they are model-dependent {\it inferred} quantities. For local 
observations this relation reduces to $\sqrt{-K}r\approx z$, which coincides with 
the SM result.

In the light-cone conformal coordinates 
the universe is infinitely old. This does not conflict with observations, e.g. that stars 
have not exhausted their nuclear fuel, because the model universe must be of finite age 
only in cosmic, not conformal, coordinates, and indeed it is (by construction). Unlike 
the SM, our model conforms with the Perfect Cosmological Principle, according to which 
there are no preferred cosmic (spacetime) events. This makes the light-cone 
conformal coordinates aesthetically appealing as well as readily amenable to 
interpretation as shown above. Clearly, the transformation $\eta=\ln(t/t_{0})$ maps 
a finite to semi-infinite time interval. This is not possible in the SM in which 
$\eta$ has an origin, $\eta=0$, since the big bang marks a genuine- rather 
than a coordinate-singularity.

{\bf Linear Perturbations and the CMB}.-
Any globally stationary (and infinitely old) universe must be found in its 
maximum entropy state. In the following, we consider entropy densities 
$\mathcal{S}$, i.e. entropy per volume of diameter $(-K)^{-1/2}$.
It is easily shown that with field mass $\sqrt{-K}$ DE saturates the holographic 
entropy (density) `bound', $\mathcal{S}_{DE}=O(\Lambda^{-1}l_{P}^{-2})=(10^{122})$. 
Second to DE is the CMB with $\mathcal{S}_{CMB}=O(10^{89})$, and indeed 
the CMB black body (BB) energy distribution is not necessarily 
indicative of a hot and dense past epoch of the universe as 
is usually thought, but rather a reflection of the requirement that in a stationary model 
the CMB should be nearly at its maximum entropy state. DM particles (if not ultralight) 
and baryons contribute much lower entropy and are therefore less constrained by entropy considerations; 
yet, their overall density distribution is rather 
homogeneous, consistent with a stationary, maximum entropy configuration.

In our model, the entire sky sphere has always been, and will always be, the observable 
causal horizon, and the observed CMB correlations on scales $\gtrsim 1^{\circ}$ are 
therefore well within our causal horizon. Thus, by construction, there is no `horizon problem'. 

The essentially {\it stationary} background cosmological model described here 
clearly has to break down at `sub-horizon' scales, as has been observed in, e.g. redshift 
evolution of galaxy metallicity and number counts 
and by the very observations of reionization itself. 
Our solution of the full system of perturbation equations [45] yields 
the gravitational potential $\phi_{\bf k}\propto\exp(\pm i\omega_{k}\eta)\times Q_{\bf k}$ with the 
dispersion relation $\omega_{k}\equiv c\sqrt{\frac{k^{2}}{3}-2K}$, and $Q_{{\bf k}}$ 
are the eigenfunctions of the Laplacian in 3D curved space. This implies that the gravitational potential 
oscillations propagate at the speed $c/\sqrt{3}$ with an effective mass $O(\hbar\sqrt{-K}/c)=
O(\hbar\sqrt{H_{0}}/c^{2})$, i.e. `horizon' scale Compton wavelength. 
Indeed, linear perturbations of 
the open and stationary background admit oscillatory rather than growing modes for 
dimensionless perturbed quantities, e.g. Newtonian gravitational potentials, and 
fractional density perturbations [45]. This {\it stability} property 
distinguishes our model from other static models, e.g. [46].

In our model there is no preferred period for the beginning of structure formation, 
and the only conceivable scenario is that, embedded in the evolving $\varphi$ phase,  
there are broken-symmetry phases of $\varphi=constant$ which locally set the standard typical 
timescale for gravitational collapse $t_{coll}=O((G\rho)^{-1/2})$.
Assuming $w_{M}=0$, the trace of the first of Eqs. (2) implies that 
a constant $\varphi$ is only possible if $R>0$ at $t\rightarrow 0$, 
i.e. the mass term in Eq. (1) is positive and $\varphi$ is constant 
by virtue of a conformal symmetry breaking. For $R>0$ the {\it static} 
equilibrium (real) solution of the last of Eqs. (4) with 
$\beta<0$ is $\varphi_{stat}=-2\sqrt{\frac{|\beta|R}{12\lambda}}
\sinh\left[\frac{1}{3}\sinh^{-1}\left(\frac{3A}{2|\beta|R}\sqrt{\frac{12\lambda}{|\beta|R}}\right)\right]$ 
where $\rho_{M}\equiv A\varphi$. In the limit $\frac{A\sqrt{\lambda}}
{(|\beta|R)^{\frac{3}{2}}}\ll 1$, 
$\varphi_{stat}\approx -\frac{A}{|\beta|R}$, an approximation justified by virtue of the facts 
that $\lambda=O(l_{P}^{2}l_{\Lambda}^{-2})$ and that CDM 
particles cannot be extremely light [45]. 
This solution is consistent with $R=O(t_{coll}^{-2})$ locally.  

Although the fundamental physical quantities may settle at 
different constant values in different gravitationally bound systems 
this leaves no detectable signatures, e.g. the {\it dimensionless} ratio 
of electromagnetic and gravitational forces is universal. 

Since in our model the universe is infinite the shape of the matter power spectrum 
is determined not by quantum fluctuations of the vacuum, but 
by an entirely different mechanism: The 
CMB power spectrum on the largest angular scales is flat due to Poisson 
noise in photon number per solid angle on the sky. The rms fractional number 
perturbation (of CMB photons) is inversely proportional to the angular scale, i.e. 
$(\frac{\Delta n}{n})_{rms}^{2}\propto\Omega^{-1}$, and therefore 
$l^{2}C_{l}^{T}\sim constant$ (where $C_{l}^{T}$ is the CMB 
angular power spectrum of temperature perturbations at multipole number $l$) 
by virtue of Parseval's identity and the fact that for 
a BB $\Delta n/n\propto \Delta T/T$.
Since both baryons and CDM contributions are linear in $\varphi$, 
which very weakly self-couples, i.e. $\lambda=O(10^{-122})$, perturbations 
are gaussian. In addition, adiabatic perturbations render 
the (infinitely old) universe globally stationary.

From the perturbation equations the observed `acoustic' oscillations 
(explained in the SM as due to plasma oscillations at recombination) 
could be understood in our model as oscillations of the 
gravitational potential on cosmological scales. 

Since in the present model the `Hubble rate', $\Gamma_{H}=O(c(-K)^{1/2})$, is at least 
$O(10^{3})$ times larger than the Compton rate, $\Gamma_{C}$; for all practical 
purposes the universe is optically thin. Therefore, the CMB temperature 
perturbation $(\Delta T/T)_{\gamma}=\delta\rho_{\gamma}/\rho_{\gamma}=O(\phi)$ 
is obtained from the collisionless Boltzmann equation. Since
$\frac{-k^{2}}{K}\phi=\delta\rho/\rho$ in the limit $k^{2}\gg -K$ [45], where $\rho$ 
is the NR energy density, 
$\eta_{M}=\eta_{B}\left(1+\frac{\rho_{CDM}}{\rho_{B}}\frac{m_{B}}{m_{CDM}}\right)$ 
is the matter-to-photon ratio, that is the ratio of matter, i.e. CDM and baryons, 
and photon number densities, where $\eta_{B}\equiv n_{B}/n_{\gamma}=
O(10^{-9})$, $\delta\rho/\rho=\delta n/n$, and $\delta n/n\propto n^{-1/2}$. 
Poisson equation with the SM-inferred $\eta_{B}$ then results in 
$M_{B}/M_{CDM}=O\left(\frac{\eta_{B}\rho_{B}}{\rho_{CDM}}(\frac{K}{k_{max}^{2}})^{2}\right)$ 
where the matter power spectrum $P(k)$ peaks at $k_{max}=O(100 Mpc)$. Adopting 
$\rho_{B}/\rho_{CDM}\approx 1/6$ deduced from observations towards bound systems, 
then yields the rough estimate of $M_{CDM}$ equals a few MeV. 
Due to the strong $M_{CDM}\propto k_{max}^{4}$ dependence 
this mass may well be a factor of few smaller or larger. 
An example for fermionic candidate in this mass range is a sterile neutrino. 
This estimate, which is a consistency requirement of our model, favors fermionic {\it light} 
DM candidates over the classical $\gtrsim 1GeV$ range for weakly interactive massive particles (WIMP's). 
Another appealing property of sterile neutrinos for an infinitely old universe is their stability.

In addition, the flat $C_{l}$ implies a flat 3D power spectrum of 
density perturbations $P(k)$ by virtue of the cosmological principle. 
On scales smaller than $\sim k_{corr}^{-1}\approx k_{max}^{-1}$ the 
density perturbations 
are not statistically independent and their correlation typically damps 
$P(k)$ by an extra $\propto k^{-4}$ factor [45] which approximately agrees 
with SM predictions, e.g. [47-49].

Since $\Gamma_{C}\ll\Gamma_{H}$ cosmic reionization at redshift $z\approx 6$ 
or higher should leave no measurable imprint on the CMB power spectra, and 
the claimed measurements of the reionization signature by analyses based on the SM 
[10, 11] consequently underestimate the cosmic contribution at `super-horizon' 
scales, e.g. that of the integrated Sachs Wolfe (SW) effect.

Since in our model all reaction rates are $\propto t^{-1}$, the optical depth to Compton 
scattering is $\tau\propto\int dt/t$, i.e. formally diverges at $t=0$, rendering the singularity 
in the energy densities practically unobservable. Although the universe is infinite, only a finite fraction 
of it is observable. We show in [45] that the observed Silk (diffusion) damping in the CMB can be explained 
as a projection effect, rather than diffusion transversal to the line of sight as in the SM.
It is also shown in [45] that CMB polarization is sourced by velocity gradients 
in the transversal direction to the line of sight. 

{\bf Summary}.-
Although the SM is an extremely useful theoretical framework 
for the description of the structure and evolution of the universe, it is based on 
certain {\it conventions} that give rise to several puzzling features. Clearly, 
energy-momentum conservation in the realm of standard GR results in a 
dynamically evolving universe, but with apparent `coincidences' that seem to single out 
the current epoch in several ways. As has been shown above, the coincidence problem 
associated with the {\it current} near equality of the energy densities of DE and matter can 
be explained away by invoking a dynamic system of natural units, i.e. scalar fields. 
Moreover, their currently observed ratio can be obtained if the sum of the canonical 
kinetic and potential energies associated with the model scalar fields is 
identified as DE, and both baryons and DM particles are described by NR fermionic fields. 
The present model applies to scenarios in which 
CDM is mainly composed of e.g. sterile neutrinos rather than WIMP's.

More generally, the framework proposed here makes use of dynamical 
$l_{P}$, $l_{C}$, $l_{\Lambda}$, etc., to effectively offset 
spacetime expansion.
From this perspective all standard `early universe' processes, e.g. 
phase transitions and production of topological defects, inflation, BBN,
recombination, etc., never take place (and indeed none of these is 
practically observable); {\it the background universe as we currently see it has always 
been the same.} 

The present model provides a {\it kinematical} alternative to the {\it dynamical} SM. 
Consequently, our model exhibits no evolution, no inflation to set up the initial 
conditions, etc. Instead, a kinematical model only requires self-consistent 
boundary conditions, e.g. $M_{CDM}$ is likely sub-$GeV$. 
The two models need only agree on dimensionless past light cone 
{\it observables}, i.e observed angular sizes, angular correlations, 
redshifts, etc., indeed a `boundary condition'. The proposed model 
explains a wide spectrum of cosmological observables in terms of self-consistency 
requirements, rather than via their {\it dynamical} evolution from `initial conditions' [45]. 

More generally, cosmology may have already provided us with a unique low energy window to the 
underlying conformal nature of gravity that we hitherto ignored by forcing our broken-phase 
solar system experience on cosmological scales. This seems to have also led to 
the misinterpretation of cosmological redshift as due to expanding
space, rather than to varying masses in the unbroken cosmic phase.

Finally, the absence of curvature singularity, and the fact that our model is infinite, 
may have far reaching implications 
on a broad spectrum of cosmological paradigms, such as quantum cosmology, the 
multiverse, chaotic inflation, and (the absence of) cosmic 
phase transitions in the very early universe, and `early universe' theories 
in general. 

{\bf Acknowledgments}.-
The author is indebted to Yoel Rephaeli for numerous constructive, critical, and 
thought-provoking discussions which were invaluable for this work.


\begin{thebibliography}{}
\bibitem{1} Starobinsky, A.~A.\ 1980, Physics Letters B, 91, 99 
\bibitem{2} Guth, A.~H.\ 1981, PRD, 23, 347 
\bibitem{3} Albrecht, A., \& Steinhardt, P.~J.\ 1982, PRL, 48, 1220 
\bibitem{4} Polnarev, A.~G.\ 1985, Soviet Astronomy, 29, 607 
\bibitem{5} Kamionkowski, M., Kosowsky, A., \& Stebbins, A.\ 1997, 
PRL, 78, 2058
\bibitem{6} Seljak, U., \& Zaldarriaga, M.\ 1997, PRL, 78, 2054 
\bibitem{7} Weinberg, S.\ 2000, arXiv:astro-ph/0005265 
\bibitem{8} Weinberg, S.\ 1989, Reviews of Modern Physics, 61, 1 
\bibitem{9} Carroll, S.~M., Press, W.~H., \& Turner, E.~L.\ 1992, ARAA, 30, 499
\bibitem{10} Hinshaw, G., Larson, D., Komatsu, E., et al.\ 2013, ApJS, 208, 19 
\bibitem{11} Planck Collaboration, Ade, P.~A.~R., Aghanim, N., et al.\ 2013, arXiv:1303.5076 
\bibitem{12} Weinberg, S.\ 1987, PRL, 59, 2607 
\bibitem{13} Efstathiou, G.\ 1995, MNRAS, 274, L73 
\bibitem{14} Efstathiou, G.\ 1995, MNRAS, 274, L73
\bibitem{15} Martel, H., Shapiro, P.~R., \& Weinberg, S.\ 1998, ApJ, 492, 29 
\bibitem{16} Peebles, P.~J.~E., \& Ratra, B.\ 1988, ApJL, 325, L17 
\bibitem{17} Wetterich, C.\ 1995, AAP, 301, 321 
\bibitem{18} Steinhardt, P.~J., Wang, L., \& Zlatev, I.\ 1999, PRD, 59, 123504 
\bibitem{19} Peebles, P.~J., \& Ratra, B.\ 2003, Reviews of Modern Physics, 75, 559
\bibitem{20} Banks, T.\ 1996, arXiv:hep-th/9601151 
\bibitem{21} Fischler, W., \& Susskind, L.\ 1998, arXiv:hep-th/9806039 
\bibitem{22} Cohen, A.~G., Kaplan, D.~B., \& Nelson, A.~E.\ 1999, PRL, 82, 4971 
\bibitem{23} Kaloper, N., \& Linde, A.\ 1999, PRD, 60, 103509 
\bibitem{24} Horava, P., \& Minic, D.\ 2000, PRL, 85, 1610 
\bibitem{25} Chiba, T., Okabe, T., \& Yamaguchi, M.\ 2000, PRD, 62, 023511
\bibitem{26} Armendariz-Picon, C., Mukhanov, V., 
\& Steinhardt, P.~J.\ 2000, PRL, 85, 4438 
\bibitem{27} Capozziello, S., Cardone, V.~F., Carloni, S., 
\& Troisi, A.\ 2003, International Journal of Modern Physics D, 12, 1969
\bibitem{28} Carroll, S.~M., Duvvuri, V., Trodden, M., 
\& Turner, M.~S.\ 2004, PRD, 70, 043528 
\bibitem{29} Rothman, T., \& Ellis, G.~F.~R.\ 1987, The Observatory, 107, 24 
\bibitem{30} Krauss, L.~M., \& Scherrer, R.~J.\ 2007, General Relativity and Gravitation, 39, 1545 
\bibitem{31} Dirac, P.~A.~M.\ 1938, Royal Society of London Proceedings Series A, 165, 199 
\bibitem{32} Canuto, V., Adams, P.~J., Hsieh, S.-H., \& Tsiang, E.\ 1977, PRD, 16, 1643 
\bibitem{33} Canuto, V., Hsieh, S.~H., \& Adams, P.~J.{\em PRL} {\bf 39} (1977) 429 
\bibitem{34} Canuto, V.~M., \& Goldman, I.\ 1983, Nature, 304, 311 
\bibitem{35} Rosen, N.\ 1982, Foundations of Physics, 12, 213 
\bibitem{36} Hwang, J.-C., \& Noh, H.\ 2002, PRD, 65, 023512 
\bibitem{37} Copeland, E.~J., Sami, M., 
\& Tsujikawa, S.\ 2006, International Journal of Modern Physics D, 15, 1753 
\bibitem{38} Deser, S.\ 1970, Annals of Physics, 59, 248 
\bibitem{39} Anderson, J.~L.\ 1971, PRD, 3, 1689 
\bibitem{40} Dabrowski, M.~P., Denkiewicz, T., \& Blaschke, D.~B.\ 2007, Annalen der Physik, 519, 237
\bibitem{41} Milne, E.~A.\ 1932, Nature, 130, 9 
\bibitem{42} Milne, E.~A.\ 1933, MNRAS, 93, 668 
\bibitem{43} Milne, E.~A.\ 1934, The Quarterly Journal of Mathematics, 5, 64
\bibitem{44} Robertson, H.~P.\ 1933, Zeitschrift für Astrophysik, 7, 153
\bibitem{45} Shimon, M.\ 2015, in preparation 
\bibitem{46} Eddington, A.~S.\ 1930, MNRAS 90, 668
\bibitem{47} Peebles, P.~J.~E.\ 1993, Principles of Physical Cosmology, Princeton University Press  
\bibitem{48} Dodelson, S.\ 2003, Modern cosmology, Academic Press, 2003  
\bibitem{49} Weinberg, S.\ 2008, Cosmology, University Press, Oxford, UK
\end{thebibliography}
\end{document}